\documentstyle[multicol,preprint,aps]{revtex}
\begin{document}
\title{A Magnetic Resonance Force Microscopy Quantum Computer with Tellurium Donors in Silicon}
\author{G.P. Berman$^1$, G.D. Doolen$^1$, and V.I. Tsifrinovich$^2$}
\address{$^1$Theoretical Division and CNLS, 
Los Alamos National Laboratory, Los Alamos, New Mexico 87545}
\address{$^2$IDS Department, Polytechnic University,
Six Metrotech Center, Brooklyn NY 11201}
\maketitle
\begin{abstract}
We propose  a magnetic resonance 
force microscopy (MRFM)-based nuclear spin quantum computer using tellurium impurities in silicon. This approach to quantum computing combines the well-developed silicon technology with expected advances in MRFM.
\end{abstract}
\newpage
\section{Introduction}
Recently, Kane \cite{1} proposed a silicon-based nuclear spin quantum computer. This proposal linked the theoretical field of quantum computation with the well-developed silicon industry technology. It was proposed in \cite{1} to use nuclear spins, $I=1/2$, of impurity phosphorus atoms ($^{31}$P) in silicon ($^{28}$Si) as qubits for quantum computation. Selective one-qubit rotation of nuclear spins can be implemented by combining the action of electrostatic gates and resonant radio frequency ({\it rf}) pulses. The electrostatic gate increases the size of the electron cloud of the selected phosphorus atom changing the hyperfine interaction between the electron spin ($S=1/2$) and the nuclear spin of the phosphorus atom. A two-qubit quantum CONTROL-NOT (CN) gate can be implemented by combining resonant {\it rf} pulses by applying two electrostatic gates acting on the neighboring phosphorus atoms. Under the action of electrostatic gates, the electron clouds of the neighboring atoms increase their size and overlap. This causes an exchange interaction between electron spins which, in turn, generates an indirect coupling between the associated nuclear spins.

To measure the state of the nuclear spin, it was proposed in \cite{1} to transfer the state of the nuclear spin to the electron spin. Then, using an  electrostatic gate, one induces a transfer of the electron from the measured phosphorus atom to the auxiliary phosphorus atom. Because of the Pauli principle, this transfer is possible only if the electron spins of the measured atom and the auxiliary atom have opposite directions.  The change of the charge of the auxiliary atom can be measured by a single-electron transistor. This attractive proposal may, however, face difficulties associated with: (a) the precise manipulation of the electron clouds using electrostatic gates, (b) the complicated diagram of transferring the nuclear spin state to the electron spin state, and (c) the application of the single-electron transistor. Vrijen et al. \cite{2} proposed a way to overcome these problems. But their proposal implements qubits in the electron spins of the phosphorus atoms. It is clear that unlike nuclear spins, electron spins cannot be isolated from their surroundings. So, the price of simplification of the original proposal \cite{1} seems to be very high. 

In our previous paper \cite{3} we proposed the MRFM quantum computer. This proposal relies on rapidly developing MRFM methods which promise single spin detection combining magnetic resonance techniques, atomic force microscopy and novel optical methods for detection of mechanical vibrations \cite{4}-\cite{6}. 
It would be very attractive to apply the idea of the MRFM quantum computer to paramagnetic impurities in silicon. However, the phosphorus atom does not fit our proposal because the large size of its electron cloud and relatively weak hyperfine interaction. 
In this paper, we propose a MRFM quantum computer based on silicon with tellurium impurities. Unlike phosphorus, a tellurium atom in silicon is a ``deep donor'' with a small size of the electron cloud and with extremely large hyperfine interaction. Application of tellurium impurities in silicon could combine advantages of MRFM with well-developed techniques of silicon technology.
ispell tellurium1.tex In section II, we discuss the design of this quantum computer. In section III, we describe quantum computation using this 
nuclear spin quantum computer. We discuss a one-qubit rotation, a two-qubit quantum CN gate, the measurement of the state of a nuclear spin, and the initialization of the nuclear spins in their ground states.
\section{MRFM Si:Te Nuclear Spin Quantum Computer}
A principal diagram of the proposed quantum computer is shown in Fig. 1. Tellurium-125 donors are placed near the surface of the silicon-28. We assume that silicon contains only $^{28}$Si non-magnetic nuclei. $^{29}$Si magnetic nuclei whose natural abundance  is 4.7\% must be eliminated. The tellurium contains $^{125}$Te nuclei, whose natural abundance is only 7\%. If a host atom in silicon is replaced by tellurium donor, two extra electrons are available. The properties of tellurium donors in silicon have been investigated elsewhere. (See, for example, \cite{7} and the references therein.) It was found that most of the implanted tellurium atoms occupy substitutional sites. The ground states of tellurium donors, as well as those of other atoms with two extra electrons, are referred as ``deep impurity levels'' in contrast  to ``shallow'' impurities like phosphorus with one extra electron whose ground state energies are of the order 50 meV. Because of the two extra electrons, tellurium donors form singly-ionized A-centers, Te$^+$, and neutral B-centers, Te$^0$. The temperature-independent  ground state energies were found in \cite{7} to be 410.8 meV for A-centers, and 198.8 meV for B-centers.

Electron spin resonance (ESR) for A-centers \cite{7} can be described by the simple spin Hamiltonian,
$$
{\cal H}=g_e\mu_B\vec B\vec S+g_n\mu_n\vec B\vec I-A\vec S\vec I,\eqno(1)
$$
where $\vec S$ is the electron spin (S=1/2) of $^{125}$Te$^+$; $\vec I$ is a nuclear spin (I=1/2) of $^{125}$Te nuclei; $\mu_B$ and $\mu_n$ are the Bohr and nuclear magnetons, $g_e$ and $g_n$ are the electron and nuclear $g$-factors: $g_e\approx 2$, $g_n\approx 0.882$; A is the constant of the isotropic hyperfine (hf) interaction, $A/2\pi \hbar\approx 3.5$ GHz. The first two terms in the Hamiltonian (1) have the same signs because the nuclear magnetic moment of $^{125}$Te is negative, as is the electron magnetic moment. For the same reason, we put in (1) a negative sign for the hyperfine interaction.

We propose to use as qubits the nuclear spins of the $^{125}$Te donors (A-centers). We assume that future advances in silicon technology will allow one to place a regular chain of $^{125}$Te donors near the surface of silicon with the distance between donors being approximately 5 nm. (See Fig. 1.) To initialize the ground states of the nuclear spins and to measure their final states, we propose using MRFM. For this purpose, the ferromagnetic particle, $P$, in Fig. 1, attached to the end of the cantilever, can move along the impurity chain selecting an appropriate tellurium ion. To implement quantum computation, we propose using the same (but non-vibrating) ferromagnetic particle which can move along the impurity chain. Next, we shall describe the operation of the proposed quantum computer.
\section{Quantum Computer Operation}
Following our proposal \cite{3}, we assume that electron spins are polarized in the positive $z$-direction. (As an example, for $B_0=10$T and at a temperature of 1K, the probability for an electron to change its direction is approximately $1.4\times 10^{-6}$.) From the other side, approximately 44\% of nuclear spins are in their excited states. To detect these nuclear spins one moves the ferromagnetic particle placed on the cantilever to a selected tellurium atom. Assuming that the distance between the ferromagnetic particle and the selected ion is 10 nm, the radius of the ferromagnetic particle is 5 nm, and the magnetic induction of the ferromagnetic particle is: $\mu_0M\approx 2.2$ T, one finds that the shift of the ESR frequency for the selected ion is, $\Delta f_e\approx 1.5$ GHz. (The corresponding magnetic field produced by the ferromagnetic particle is approximately $5.4\times 10^{-2}$ T \cite{3}.) The ``natural'' ESR frequency for $B_0=10$ T is, $f_e\approx 280$ GHz. The hyperfine shift of the ESR frequency is, $f_{hf}=A/4\pi\hbar\approx 1.75$ GHz \cite{7}. For an ion, the magnetic dipole field produced by its two neighbor electron spins was estimated as $1.5\times 10^{-5}$ T. The magnetic dipole field produced by all other electron spins does not exceed $3\times 10^{-6}$ T \cite{3}. The corresponding shifts of the ESR frequency are, $f_{ed}\approx 0.42$ MHz and $f^\prime_{ed}<0.08$ MHz. We assume that the amplitude of the {\it rf} pulse, $B_1$, in frequency units (the Rabi frequency) is greater than $f_{ed}$. So, the dipole contribution to the ESR frequency can be ignored. Thus, applying the {\it rf} pulses with the frequency,
$$
f\approx f_e+f_{hf}+\Delta f_e,\eqno(2)
$$
one induces oscillations of the electron spin of a selective tellurium ion only if the nuclear spin of the ion is in its ground state. The oscillating electron spin, in turn, induces resonant vibrations of the cantilever which can be detected by MRFM methods. A discussion on modified MRFM techniques for detection of a single electron spin and related estimates for the MRFM quantum computer can be found in our previous papers \cite{3,8}.

Thus, tellurium atoms detected by MRFM have their nuclear spins in their ground state. To drive other nuclear spins to their ground states, one moves the non-vibrating ferromagnetic particle to a selected tellurium atom whose nuclear spin is in the excited state. The ``natural'' NMR frequency for $^{125}$Te  nuclear spin in an external magnetic field of 10 T is 134.5 MHz. The hyperfine ``shift'', $f_{hf}\approx 1.75$ GHz is larger than the ``natural'' frequency. The additional shift caused by the magnetic field of the ferromagnetic particle is, $\Delta f_n\approx 0.73$ MHz. Applying an {\it rf} $\pi$-pulse with  frequency,
$$
f=f_n+f_{hf}+\Delta f_n-f_{nd}-f^\prime_{nd},\eqno(3)
$$
one drives the nuclear spin into its ground state.

In Eq. (3), the frequency, $f_{nd}$, is the NMR shift caused by the dipole field of the electron spins of the neighbor ions, and $f^\prime_{nd}$ is caused by  electron spins of all other ions. For $^{125}$Te, the frequency $f_{nd}\approx 200$ Hz, and $f^\prime_{nd}< 40$ Hz. Applying an {\it rf} pulse with a nuclear Rabi frequency larger than $f_{nd}$, one can neglect the dipole contribution. The same method can be used to implement a one-qubit rotation. To implement a two-qubit gate, we propose using the magnetic dipole interaction between electron spins of tellurium ions. For this purpose, one moves the non-vibrating ferromagnetic particle to a tellurium ion containing a control nuclear spin (a control qubit). Then, one applies an {\it rf} pulse with frequency, $f=f_e+f_{hf}+\Delta f_e$. This pulse drives the electron spin into its excited state if the control nuclear spin is in the ground state. Next, one moves the non-vibrating ferromagnetic particle to the tellurium ion containing the target nuclear spin (a target qubit). Now, it is important to use a selective {\it rf} pulse whose frequency is,
$$
f=f_n+f_{hf}+\Delta f_n-f^\prime_{nd},\eqno(4)
$$
and whose Rabi frequency is less than 200 Hz. This pulse changes the state of the target nuclear spin if the dipole contribution from neighbor electron spins
cancels out: $f_{nd}=0$. It happens only if the control nuclear spin was in the ground state. Finally, one moves the non-vibrating ferromagnetic particle back to the ion containing the control nuclear spin and applies a $\pi$-pulse with the frequency (2), to return the electron spin in the ground state (if it was in the excited state). Thus, three {\it rf} pulses together implement an ``inverse'' quantum CN gate: the target qubit changes its state if the control qubit is in the ground state. to implement the ``standard'' quantum CN gate one can apply a {\it rf} $\pi$-pulse with the frequency (3). In this case, the target nuclear spin changes its direction if the electron spin of the neighbor ion did not change its state, i.e. if the control nuclear spin was in the excited state. The final measurement of the nuclear state can be implemented using  MRFM in the same way as the measurement of the initial nuclear states.
\section{Conclusion}
We describe a MRFM nuclear spin quantum computer using tellurium-125 singly ionized donors placed near the surface of silicon-28. Our proposal relies on the expected advances in MRFM which promises the detection of a single electron and it relies on further developments in silicon technology.\\ \ \\     
\quad\\
{\large\bf Acknowledgments}\\ \ \\
We thank P.C. Hammel for valuable discussions. This work  was supported by the Department of Energy under contract W-7405-ENG-36 and by the National Security Agency.
\newpage
\newpage
\quad\\
{\Large\bf Figure captions}\\ \ \\
Fig. 1:~A diagram of the proposed quantum computer. The circles indicate $^{125}Te^+$ ions implanted near the surface of the silicon substrate; $I$ is the nuclear spin; $S$ is the electron spin (electron spins are shown in their ground states); $P$ is the ferromagnetic particle (vibrating or non-vibrating); $d$ =15 nm; $a$=5 nm.

\end{document}